\documentclass[twocolumn,superscriptaddress,notitlepage,aps,pra,showkeys,10pt]{revtex4-1}

\usepackage{graphicx}% Include figure files
\usepackage{dcolumn}% Align table columns on decimal point
\usepackage{bm}% bold math
\usepackage{amssymb}
\usepackage{epsfig}
\begin{document}
\title{Tunnel Spectroscopy of a Proximity Josephson Junction}
\author{M. Meschke}
\affiliation{Low Temperature Laboratory, Aalto University, P.O. Box 13500, FI-00076 Aalto, Finland}
\email{matthias.meschke@aalto.fi}
\author{J. T. Peltonen}
\affiliation{Low Temperature Laboratory, Aalto University, P.O. Box 13500, FI-00076 Aalto, Finland}
\author{J. P. Pekola}
\affiliation{Low Temperature Laboratory, Aalto University, P.O. Box 13500, FI-00076 Aalto, Finland}
\author{F. Giazotto}
\affiliation{NEST Istituto Nanoscienze-CNR and Scuola Normale Superiore, I-56127 Pisa, Italy}

%\date{\today}
%\pacs{72.25.-b,85.75.-d,74.50.+r}

\begin{abstract}

We present tunnel spectroscopy experiments on the proximity effect in lateral superconductor-normal metal-superconductor (SNS) Josephson junctions. Our weak link is embedded into a superconducting (S) ring allowing phase biasing of the Josephson junction by an external magnetic field. 
We explore the temperature and phase dependence of both the induced mini-gap and the modification of the density of states in the normal (N) metal. Our results agree with a model based on the quasiclassical theory in the diffusive limit. The device presents an advanced version of the superconducting quantum interference proximity transistor (SQUIPT), now reaching flux sensitivities of 3 nA$/\Phi_0$ where $\Phi_0$ is the flux quantum.

\end{abstract}

\maketitle

\section{Introduction}
\label{sec:intro}

Proximity effect \cite{degennes} appears when superconducting correlations penetrate through a clean boundary into a normal-type conductor \cite{heersche,jherrero,keizer,morpurgo,kasumov,cleuziou,doh,xiang,pothier,courtois,morpurgo1,giazotto,baselmans}.
As a consequence, the local density of states (DOS) is modified in the normal metal, and a mini-gap is induced whose size can be controlled by changing the macroscopic phase of the superconducting order parameter across the weak link \cite{belzig2,zhou,cuevas,gueron,belzig,petrashov,petrashov2,sueur,pillet}. 

Here we report an experimental study of a proximized copper metal island with a length $L$ on the order of the superconducting coherence length using tunnel spectroscopy. A complementary study of such a system uses scanning tunnel microscopy and allows space resolution of the proximized system consisting of a silver weak link \cite{sueur}. In comparison, our approach benefits from a better control of the tunnel probe with respect to the matching of the tunnel junction impedance to the amplifier yielding superior signal to noise ratio. We achieve an enhanced energy resolution and we can design a device with well defined properties for a high field sensitivity. 

The measurements agree qualitatively with theoretical predictions of the phase dependent mini-gap and reveal unambiguously the predicted sharp drop of the DOS \cite{cuevas} in the normal metal at energies corresponding to the gap edge of the superconductor. Our modeling is further optimized when we take the influence of the electromagnetic environment surrounding our setup into account \cite{ingold}. We can observe a robust feature even at elevated temperatures despite the small magnitude of the mini-gap with respect to the gap of the BCS superconductor. Finally, our result demonstrates a ten-fold improvement of the performance of the superconducting quantum interference proximity transistor (SQUIPT) \cite{giazotto2}. We achieve flux sensitivities well below $ 10^{-5}\Phi_0$Hz$^{-1/2}$ for the present design, still limited by the amplifier noise, with an intrinsic power dissipation ($\sim 100$ fW) which is several orders of magnitude smaller than in conventional superconducting interferometers \cite{clarke,tinkham,likharev}.

\begin{figure}[t!]
\includegraphics[width=8.3cm]{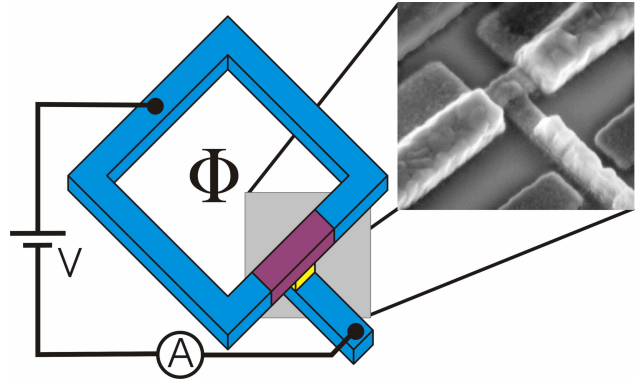}
\caption{(Color online) Schematic view of the experimental setup: a superconducting Al loop with an area of $100\,\mu$m$^2$  (blue) is interrupted using a weak link made out of copper (magenta). One superconducting tunnel probe is attached to the middle of the weak link. The scale-up image on the right depicts a scanning electron micrograph of the sample core showing the Al electrode (of width $\sim 100$ nm) connected via a tunnel junction to the Cu wire as well as the Al/Cu/Al superconductor-normal metal-superconductor proximity junction with a length $L=300$ nm. $\Phi$ symbolizes the applied magnetic flux threading the loop. Our basic electrical setup consists of a voltage bias and current measurement using a preamplifier at room temperature.  
}
\label{Figure1_setup}
\end{figure}

\section{Device characteristics}
\label{sec:device}

The device shown in Fig. \ref{Figure1_setup} is fabricated using electron-beam lithography and three angle shadow evaporation. At first, a 15-nm-thick aluminum (Al) layer is deposited and oxidized for 5 minutes with oxygen pressure of 5 mbar to form the tunnel barrier. 
Right after, typically 20 nm of copper (Cu) is deposited to form the weak link connected to the normal metal-insulator-superconductor (NIS) probe. The resulting tunnel resistance is of the order of $\sim 50$ k$\Omega$ for a junction with size of $\approx$ 100 x 100 nm$^2$. 
Finally, the superconducting Al loop with an area of $\simeq 100\,\mu$m$^2$ is placed on top, forming the clean contacts to the copper island, where the normal metal region extends laterally for about 200 nm underneath the 200-nm-thick Al leads. This sequence of metal deposition allows a thick superconducting layer forming the loop, thus minimizing the influence of the inverse proximity effect in the ring material as well as lowering its self-inductance. Moreover, in this way a reduction of the weak link thickness is only limited by the grain size of the Al film. 

In our system, proximity effect is strong as the weak link length is of the order of the superconducting coherence length, and the tunnel junction allows to probe the density of states in the proximized region. The ring geometry allows us to change the phase difference across the normal metal-superconductor boundaries through the modulation of an external magnetic field which gives rise to flux $\Phi$ through the loop area. This modifies the DOS in the normal metal, and hence the transport through the tunnel junction.

\begin{figure}[t!]
\includegraphics[width=4.0cm]{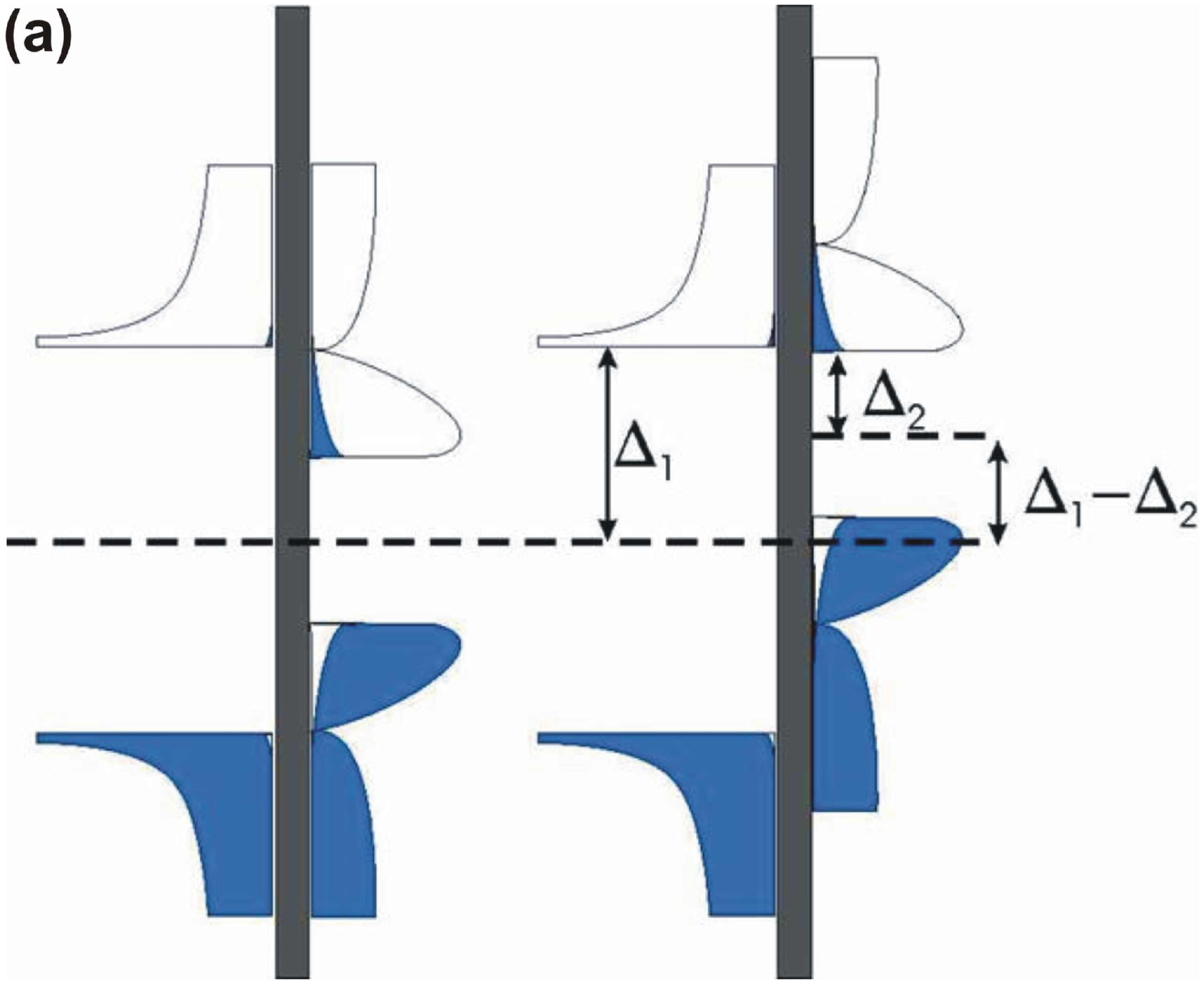}
\includegraphics[width=4.5cm]{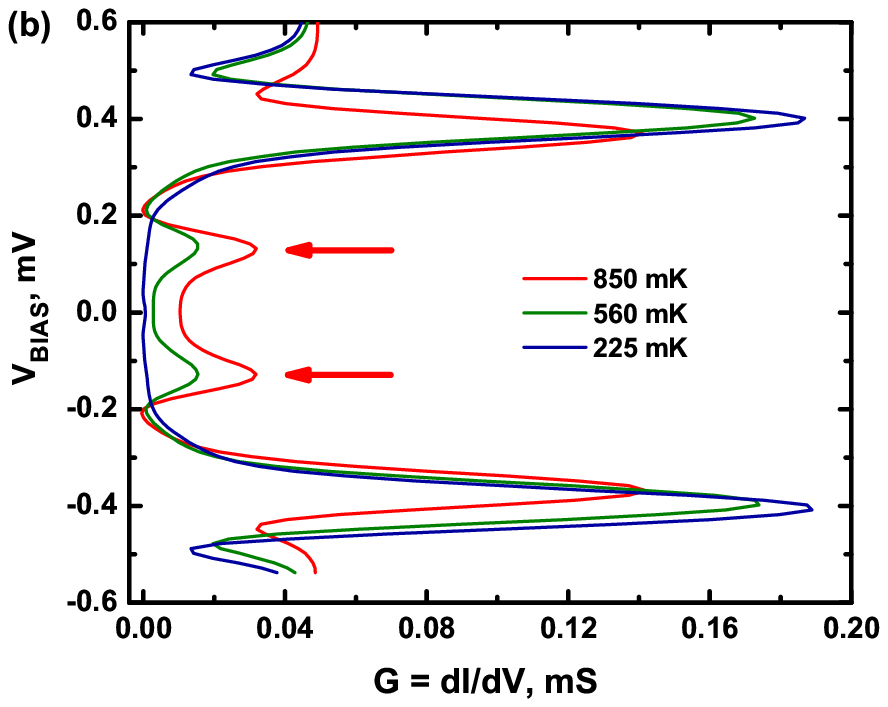}
\includegraphics[width=8.3cm]{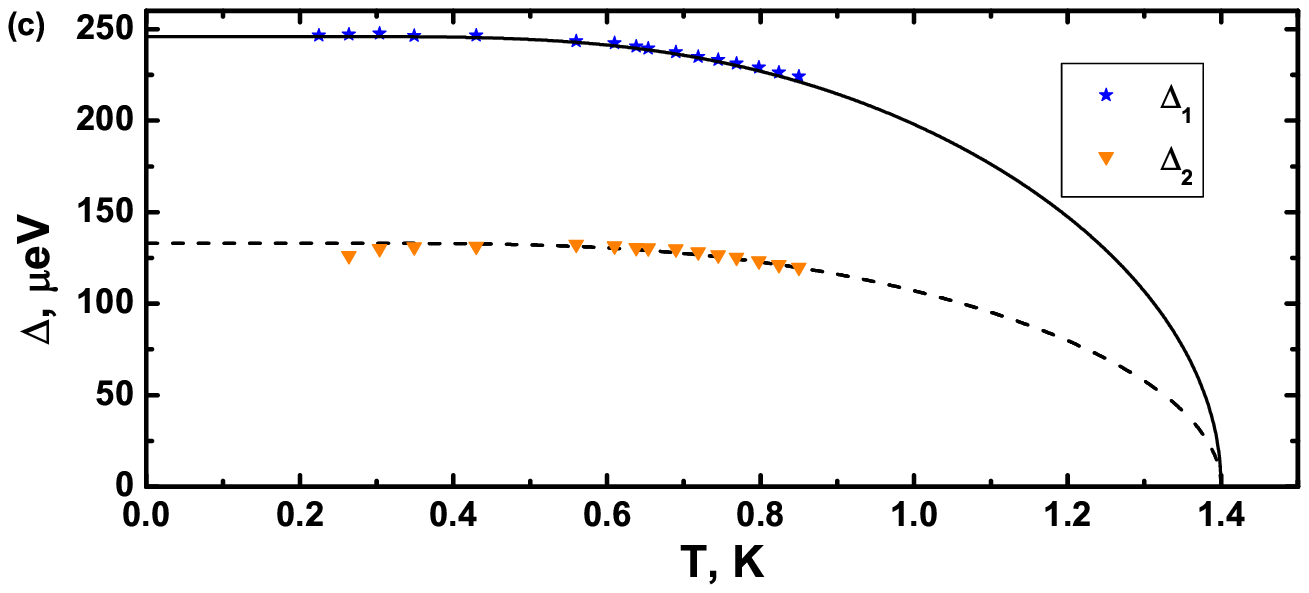}
\caption{(Color online) (a) Sketch of the density of states of a superconductor (Al, left of the tunnel junction) and proximimized Cu island (right of the tunnel junction) at zero bias (left) and at bias voltage of $(\Delta_1$-$\Delta_2)/e$ and the occupation at elevated temperature of $T \approx 1/3 T_{\rm{C}}$ of Al. Blue areas symbolize occupied states. (b) Measured differential conductance versus voltage
curves at three different bath temperatures. Arrows indicate the position of $\Delta_1$-$\Delta_2$. (c) Measured aluminum gap (blue stars) and induced mini-gap (orange triangles) as a function of bath temperature $T$. The BCS gap temperature dependence of aluminum is shown by the full line calculated using the measured gap value. The dashed line follows the same result scaled down by the ratio of the two gaps ($\Delta_2 / \Delta_1$). }
\label{Figure2_minigap}
\end{figure}

Figure \ref{Figure2_minigap}(a) depicts a sketch of the DOS of the superconducting probe junction attached to the middle of the weak link: the BCS-like density of states in the superconductor acts as an energy filter due to the superconducting gap ($\Delta_1$). 
Only few quasiparticles are excited at temperatures well below the critical temperature $T_{\rm{C}}$ of Al. 
The DOS in the proximized system on the other side of the barrier is characterized by three main features \cite{cuevas}: 
(i), a mini-gap ($\Delta_2$) with  magnitude of about one half of the Al gap ($\Delta_1$) for the dimensions of our sample; 
(ii), a sharp drop of the density of states to zero at energy values corresponding to the divergence at the gap edge of the Al superconductor; (iii), a noticeable number of excited quasiparticles already at bath temperatures ($T$) around 1/3 $T_{\rm{C}}$ of Al due to the smaller gap value. 
The curves in Fig. \ref{Figure2_minigap}(b) show the differential conductance measured with an applied voltage bias sweep plus a small voltage modulation superimposed (20 $\mu\rm{V}$) at $\Phi=0$. Then the current is measured using a preamplifier and a lock-in amplifier. 
All described features of the proximized system can be clearly observed at an elevated temperature of 550 mK and 850 mK. First, marked with red arrows, the mini-gap increases the conductance at a bias voltage of $\approx$ 150 $\mu$V when the edge of the mini-gap and the aluminum BCS gap are aligned at $V_{\rm{BIAS}} = (\Delta_1-\Delta_2)/e$. Second, the largest conductance peak appears when the BCS gap edge faces the states at the mini-gap edge at 350 $\mu$V at $V_{\rm{BIAS}} = (\Delta_1+\Delta_2)/e$. Finally, a sharp drop of conductance is observed at $V_{\rm{BIAS}} = 2\Delta_1/e \simeq 500\ \mu V$ when the DOS in the proximized metal facing the divergence in the BCS system is zero. The observed features are consistent with an aluminum gap of $\sim 246$ $\mu$eV and a mini-gap size of $\sim 130$ $\mu$eV. 

The mini-gap is still observable at 850 mK, and its magnitude reduces only slightly by increasing the temperature up to 850 mK. 
A BCS superconductor with a gap equal to the mini-gap in the proximized region [i.e., $\Delta(0) = \Delta_2$ = 130 $\mu$eV] would exhibit a critical temperature of $T_{\rm{C}} \simeq \Delta(0)/(1.764k_{\rm{B}}) \approx$ 850 mK, where $k_B$ is the Boltzmann constant. 
By contrast, we observe that the temperature dependence of $\Delta_2$ follows closely that of the aluminum gap, scaled down by almost a factor of two [see Fig. \ref{Figure2_minigap}(c)].

Lowering the temperature changes the picture as the peak marking the alignment of the mini-gap with the Al gap vanishes. This is because the quasiparticle number becomes negligible in the proximized layer. At the same time, the BCS aluminum gap slightly increases, approaching the zero-temperature value.  

\begin{figure}[t!]
\includegraphics[width=8.5cm]{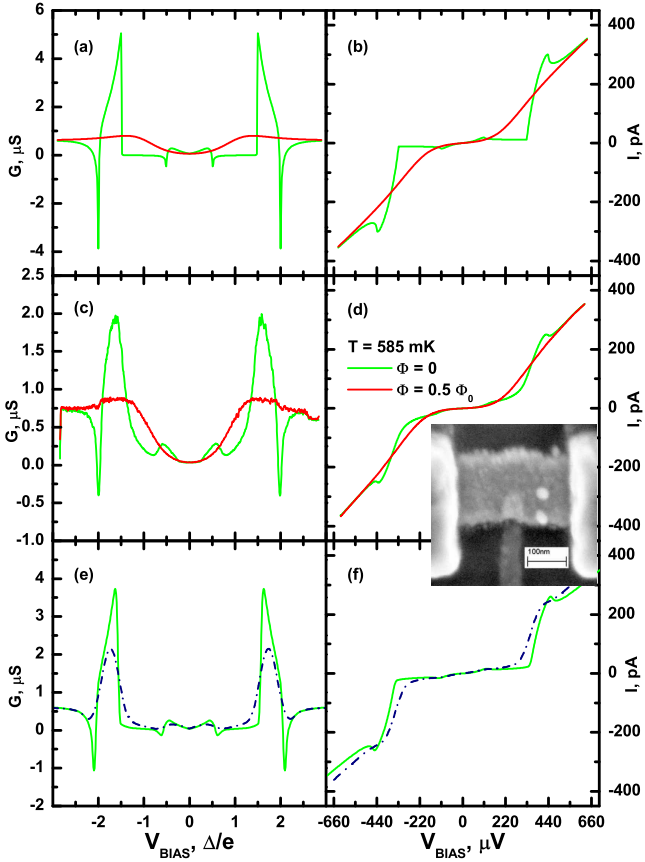}
\caption{(Color online) Conductance (a),(c),(e) and current (b),(d),(f) as a function of applied bias voltage. Measurements (c),(d) are compared to the theoretical models (a),(b) and (e),(f). The electron micrograph depicts the investigated SNS junction with the attached tunnel probe.
(top) Calculated current vs voltage curve (b) assuming $\Delta_1(0) = 220 \ \mu \rm{eV}$ and weak link length $L = 2 \xi _0$ and the corresponding conductance curve (a) at two extreme flux values where the mini-gap is maximized ($\Phi=0$, green) and fully closed ($\Phi=\Phi_{0}/2$, red).
(middle) Measured currents at a bath temperature of 585 mK (d) and conductances (c) calculated as numerical derivative of the current as a function of applied voltage for the same flux values as above. 
(bottom) Current vs voltage curve (f) and corresponding conductance (e) calculated including the influence of an electromagnetic environment at $\Phi=0$ with $T_{env}=4.2$ K (dashed blue line) and $T_{env}=T=585$ mK (green line). 
}
\label{Figure3_IV}
\end{figure}

\section{Theoretical description of the proximized system}
\label{sec:prox}

The proximity effect in the diffusive N island can be described with the Usadel equations \cite{usadel} 
which can be written as \cite{usadel,belzig}
\begin{eqnarray}
\hbar D\partial^2_x\theta =-2i\varepsilon\sinh(\theta) +\frac{\hbar D}{2}
\left(\partial_x\chi \right)^2\sinh(2\theta) \nonumber\\
\text{sinh}(2\theta)\partial_x \theta\partial_x \chi+\text{sinh}^2(\theta)\partial_x^2\chi=0,
\label{retard}
\end{eqnarray} 
where
$D$  is the diffusion coefficient and the energy $\varepsilon$ is relative to the chemical potential in the superconductors.
$\theta$ and $\chi$ are complex scalar functions of position $(x)$ and energy.
For perfectly transmitting interfaces the boundary conditions at the NS contacts reduce to $\theta(\pm L/2)=\mbox{arctanh}(\Delta_1/\varepsilon)$ and $\chi(\pm L/2)=\pm\varphi/2$,
where $\varphi$ is the phase difference across the SN boundaries. 
For lower-transparency SN interfaces the proximity effect in the wire will be reduced thus weakening the effects described below \cite{belzig}. 
Moreover, we choose a step-function form for the order parameter, i.e., constant in S and zero in the N wire, 
and we assume the BCS temperature dependence of $\Delta_1(T)$ with critical temperature $T_{\text{C}}=\Delta_1(0)/(1.764k_{\text{B}})$, where $\Delta_1(0)$ is the zero-temperature order parameter.
The DOS in the N region normalized to the DOS at the Fermi level in the absence of proximity effect is then given by $\mathcal{N}_N(x,\varepsilon,\varphi)= \mbox{Re}\left\{\cosh\left[\theta(x,\varepsilon,\varphi)\right]\right\}$. 

For a comparison of the above theory with the experiment we have to take into account the influence of single-electron coulombic effects on charge  transport \cite{ingold}.
The  quasiparticle current (evaluated in the middle of the wire, i.e., $x=0$) through the tunnel junction biased at voltage $V$ can be written as \cite{ingold,wolf}
\begin{eqnarray}
I(V,\varphi)=\frac{1}{eR_t}\int_{-\infty}^{\infty} d\varepsilon \mathcal{N}_N (\varepsilon,\varphi)\int_{-\infty}^{\infty} dE P(E,T_{env})\\\nonumber
\times [f_0(\varepsilon)\mathcal{F}_1 (\varepsilon,E,V)-(1-f_0(\varepsilon))\mathcal{F}_2 (\varepsilon,E,V)],
\label{current}
\end{eqnarray} 
where $\mathcal{F}_1(\varepsilon,E,V)=\mathcal{N}_S(\varepsilon+eV-E)[1-f_0(\varepsilon+eV-E)]$, $\mathcal{F}_2(\varepsilon,E,V)=\mathcal{N}_S(\varepsilon+eV+E)f_0(\varepsilon+eV+E)$,
$\mathcal{N}_S(\varepsilon,T)=|\varepsilon|/\sqrt{\varepsilon^2-\Delta_1(T)^2}\Theta[\varepsilon^2-\Delta_1(T)^2]$ is the normalized DOS of the S probe, $\Theta(y)$ is the Heaviside step function, $f_0(\varepsilon)=[1+\text{exp}(\varepsilon/k_BT)]^{-1}$ is the Fermi-Dirac energy distribution at temperature $T$, and $e$ is the electron charge. 
Furthermore, $P(E,T_{env})=\frac{1}{2\pi \hbar}\int_{-\infty}^{\infty} dt e^{J(t,T_{env})+iEt/\hbar}$ is the probability for the electromagnetic environment to absorb energy $E$ in a tunneling event \cite{devoret}, $J(t,T_{env})=\frac{1}{R_K}\int_{-\infty}^{\infty} \frac{d\omega}{\omega}\text{Re}[Z(\omega)](e^{-i\omega t}-1)[1+\text{coth}(\frac{\hbar \omega}{2 k_B T_{env}})]$ is the phase correlation function, $Z(\omega)=R_{env}(1+i\omega R_{env}C)^{-1}$ is the impedance of a purely resistive environment, $C$ is the junction capacitance, $R_{env}$ and $T_{env}$ are the resistance and temperature of the environment, respectively, and $R_K\simeq 25.8$ k$\Omega$ is the Klitzing resistance.
By neglecting the ring inductance the phase difference across the N wire becomes $\varphi=2\pi\Phi/\Phi_0$, where $\Phi$ is the total flux through the loop area, and $\Phi_0=2.067\times 10^{-15}$ Wb is the flux quantum.

Data shown in Fig. \ref{Figure3_IV}(c) and (d) are for a sample using a tunnel probe with further reduced size of $\approx$ 40 x 40 nm$^2$ and consequently enhanced resistance of the order of $\sim 1$ M$\Omega$ (see the inset of Fig. 3). Any influence of the superconducting probe on the density of states in the island is therefore effectively suppressed and negligible in comparison to the dominant influence of the two clean contacts. 
Measurements were performed at 585 mK as the quasiparticle population within the proximized system is at this elevated temperature sufficient to reveal its structure [see Fig. \ref{Figure2_minigap}(b)]. 
The current vs voltage curve and differential conductance show the typical behavior of a NIS junction at $\Phi=\Phi_0/2$ (red curve) with a very low conductance within the gap, enhanced conductance at the gap edges and thereafter lowering conductance toward the asymptotic values at high voltages. For $\Phi=0$ the mini-gap is visible and a sharp feature occurs at 2$\Delta_1/e$. 
Figure \ref{Figure3_IV}(b) shows the current from Eq. (2) and the corresponding conductance, $G=dI/dV$, (a) at two extreme flux values where the mini-gap is maximized ($\Phi=0$) and fully closed ($\Phi=\Phi_0/2$) assuming no influence of the electromagnetic environment (i.e., $R_{env}=0$). 
As representative parameters for the present sample we set $\Delta_1(0)= 220\,\mu$eV, $D=0.01$ m$^2$s$^{-1}$ and $L=2\xi_0\simeq 350$ nm, where $\xi_0=\sqrt{\hbar D/\Delta_1(0)}$ is the coherence length.  
Resemblance between experiment and theory is evident, the latter predicting the overall shape, although the experimental data appear somewhat smeared with respect to  prediction.  
Finite influence of the environment can account for the observed broadening. 
For illustration, Fig. \ref{Figure3_IV}(f) and (e) show the current and conductance, respectively, calculated at $\Phi=0$ assuming $R_{env}=250\,\Omega$ and $C=0.1$ fF for two different values of $T_{env}$.  
In particular, a good qualitative agreement with the experiment is obtained by setting $T_{env}=T=585$ mK.

\section{SQUIPT performance}
\label{sec:squipt}

\begin{figure}[t!]
\includegraphics[width=5.15cm]{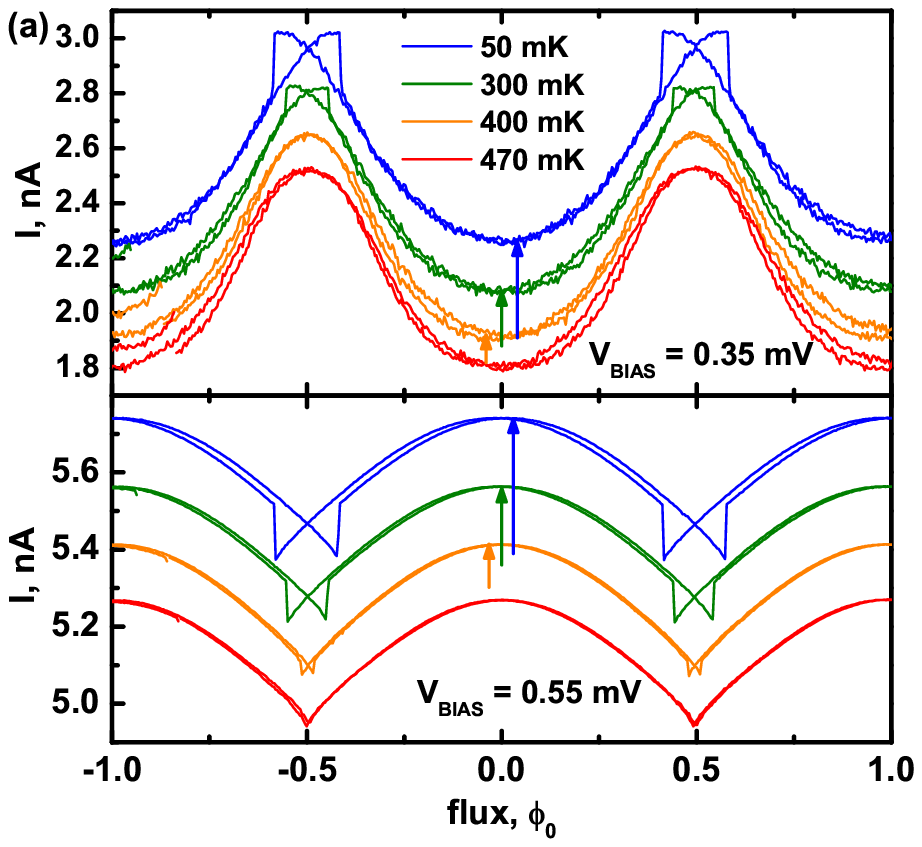}
\includegraphics[width=3.35cm]{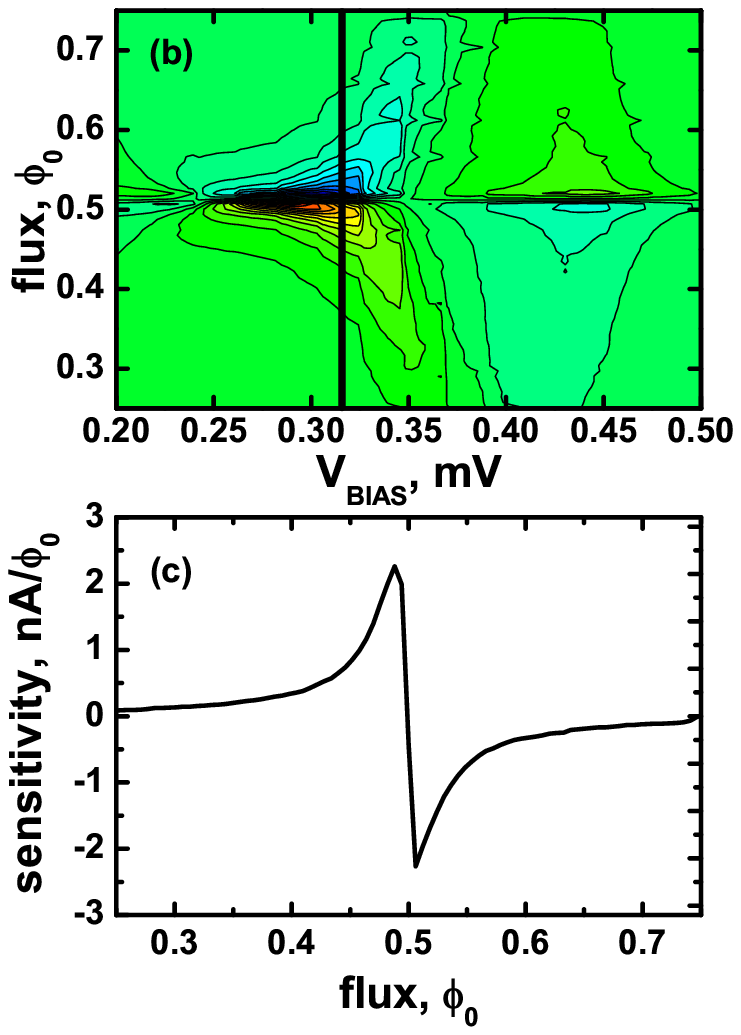}
\caption{(Color online) SQUIPT. (a) Current through the device at two different voltage bias ($V_{\rm{BIAS}}$) points as a function of magnetic flux through the ring. Curves corresponding to different temperatures are vertically offset for clarity by current values indicated with the arrows. (b) Contour plot of $\partial I/\partial\Phi$ measured with lock-in technique using a field modulation as a function of voltage bias and external flux. Color scale is from -3 $\rm {nA/\Phi_0}$ (blue) to 3 $\rm {nA/\Phi_0}$ (red) with a step of 0.3 $\rm {nA/\Phi_0}$. (c) Vertical line in the top panel corresponds to $V_{\rm{BIAS}}=0.32$ mV plotted here and shows the SQUIPT sensitivity $\partial I/\partial\Phi$ at this biasing point as a function of  $\Phi$.}
\label{Figure4_SQUIPT}
\end{figure}

In this paragraph, we discuss the performance of the present improved design of a SQUIPT \cite{giazotto2}. 
Figure \ref{Figure4_SQUIPT}(a) shows the measured current through the device, modulated as a function of flux piercing the ring, when the NIS junction is biased at a constant voltage. 
Current biasing the device and using a voltage readout represents a complementary setup with similar properties. 
The modulation amplitude shows only a weak temperature dependence reflecting the constant gap magnitudes, when the temperature stays below $\simeq$ 500 mK ($\approx$ 1/3 $T_{\rm{C}}$ of aluminum). 
Hysteresis appears towards low temperatures as soon as the Josephson inductance of the weak link $L_{\rm{J}}=\Phi_0/(2\pi I_{\rm{C}})$, where $I_{\rm{C}}$ is the critical current of the SNS junction, falls below the self-inductance of about 8 pH of the superconducting ring. We expect \cite{sns} a value of $I_{\rm{C}}$ on the order of $500\ \rm {\mu A}$ toward zero temperature corresponding to a Josephson inductance on the order of 1 pH. 
$I_{\rm{C}}$ can be reduced either by increasing $T$ or by shrinking the weak-link cross section \cite{dubos, heikkila} which leads to an increased normal-state resistance, therefore tuning these two parameters can eliminate this undesired effect.

Panel (b) of Fig. 4 displays the SQUIPT sensitivity ($\partial I/\partial\Phi$) as a function of external flux and bias voltage, measured using a magnetic field modulation amplitude  of 20 $\mu \Phi_0$ at a frequency of a few tens of Hz and a lock-in amplifier with a room temperature current preamplifier. The device sensitivity reaches $\sim 3$ nA/$\Phi_0$ (see Fig. 4(c)) in the presented voltage biased scheme and around $\sim 1.5\ \rm {m V/\Phi_0}$ in a current biased setup. 
These improved figures are the direct consequence of the shortened weak link and therefore enhanced magnitude of the induced mini-gap with respect to the earlier work \cite{giazotto2}. 
This corresponds to a flux resolution of $\simeq 2 \times 10^{-6}\ \Phi_0/\sqrt{\rm{Hz}}$ when using a typical room temperature low noise current amplifier with a specified noise level of 5 fA/$\sqrt{\rm{Hz}}$. 
In our experiment, we observe three-fold higher noise levels, still resulting in flux resolution of $6 \times 10^{-6}\Phi_0 /\sqrt{\rm{Hz}}$ at 1 kHz. 
The corresponding figures of merit using voltage measurements are similar: a typical noise level of 8 nV/$\sqrt{\rm{Hz}}$ at 1 kHz yields flux resolution of $ \approx 6 \times 10^{-6}\ \rm {\Phi_0}/ \sqrt{\rm{Hz}}$. 
We note that these figures are still determined by the noise of the amplifiers and not by the device. The observed performance can be reached as the tunnel junction impedance  ($\sim 200$ k$\Omega$) is well matched to the working point where the amplifier shows a minimum noise.

With a typical 1 nA of current output at $V_{\rm{BIAS}}\simeq 100\,\mu$V we get a total dissipated power ($P$) in the SQUIPT which is of the order of $P\sim 100$ fW. Such a power is several orders of magnitude smaller than in conventional superconducting quantum interference devices (SQUIDs) \cite{clarke,tinkham,likharev}.  
A suppressed $P$ value is moreover beneficial in order to prevent substantial electron heating in the N region. 
At low temperature (typically below 1 K) the main contribution to heat relaxation in the island is related to electron-acoustic phonon interaction \cite{rmp} ($\dot{Q}_{e-ph}$)  which, including proximity effect, can be approximated at $\Phi=0$ as \cite{HG} $\dot{Q}_{e-ph}\simeq \dot{Q}_{e-ph}^Ne^{-3.7E_{Th}/k_BT_e}$, where $E_{Th}=\hbar D/L^2$ is the Thouless energy of the SNS junction,
$\dot{Q}_{e-ph}^N=\Sigma\mathcal{V}(T_{e}^5-T^5)$ is the heat flux in the normal state, $\Sigma=2\times 10^9$ Wm$^{-3}$K$^{-5}$  is the electron-phonon coupling constant in Cu,\cite{rmp} $\mathcal{V}\simeq 2.8\times 10^{-21}$ m$^3$ is the island volume and $T_e$ is the electronic temperature in N. Under continuous power injection $P$ the steady-state $T_e$ follows from the solution of the energy-balance equation $P+\dot{Q}_{e-ph}=0$ which would give $T_e \simeq 650$ mK at $T=500$ mK in our structures, i.e., a temperature for which the electronic properties of the N layer are still quite similar to those at lower bath temperature [see Fig. 2(c)].
In the above discussion we supposed the Al superconducting loop to act as an ideal Andreev mirror for the heat flux, although at these bath temperatures the thermal conductance of the S ring is already sizable to provide thermalization of electrons in the Cu strip \cite{timo}. This latter statement is additionally supported by our experimental observation of a growing hysteresis when lowering the bath temperature from 300 mK to 50 mK [see Fig. \ref{Figure4_SQUIPT}(a)]. 
%As a consequence, electron thermalization, at least down to 300 mK, is guaranteed in our system by the thermal conductance of the superconducting loop. 

\section{Conclusion}
\label{sec:conc}

To summarize, we investigated a proximized system using tunnel spectroscopy. Our experimental findings are well described using a standard treatment based on quasiclassical theory of superconductivity in the diffusive limit expanded by the influence of the electromagnetic environment of our device. The approach has sufficient resolution to study the alterations within the DOS of a proximized normal metal. We show that the device has the potential for realizing flux sensors which combine very low power dissipation and a simple setup with a competitive sensitivity. A strength of the concept is that the device can be optimized by adjusting the size and impedance of the tunnel junction independently of the weak link. 

\section{Acknowledgements}
%We gratefully acknowledge N.N. for fruitful discussions. 
The work was partially supported by the European Community's FP7 Programme under Grant Agreement No. 228464 (MICROKELVIN, Capacities Specific Programme) and by the Academy of Finland under project number 139172.

%
%\bibitem{garcia}
%C. Pascual Garc\'{i}a and F. Giazotto, Appl. Phys. Lett. \textbf{94}, 132508 (2009).
%\bibitem{rmp}
%F. Giazotto, T. T. Heikkil$\ddot{\text{a}}$, A. Luukanen, A. M. Savin, and J. P. Pekola, Rev. Mod. Phys. \textbf{78}, %217 (2006).


\begin{thebibliography}{99}


\bibitem{degennes}
P. G. de Gennes, {\em Superconductivity of Metals and Alloys} (W. A. Benjamin, New York, 1966).
\bibitem{heersche}
H. B. Heersche, P. Jarillo-Herrero, J. B. Oostinga, L. M. K. Vandersypen, and A. F. Morpurgo, Nature \textbf{446}, 56 (2007).
\bibitem{jherrero}
P. Jarillo-Herrero, J. A. van Dam, and L. P. Kouwenhoven, Nature \textbf{439}, 953 (2006).
\bibitem{keizer}
R. S. Keizer \emph{et al.}, Nature \textbf{439}, 825 (2006).
\bibitem{morpurgo}
A. F. Morpurgo, J. Kong, C. M. Marcus, and H. Dai, Science \textbf{286}, 263 (1999).
\bibitem{kasumov}
A. Yu. Kasumov, M. Kociak, S. Gu$\acute{\text{e}}$ron, B. Reulet, V. T. Volkov, D. V. Klinov, and H. Bouchiat, Science \textbf{291}, 280 (2001).
\bibitem{cleuziou}
J.-P. Cleuziou, W. Wernsdorfer, V. Bouchiat, T. Ondarcuhu, and M. Monthioux, Nature Nanotech. \textbf{1}, 53 (2006).
\bibitem{doh}
Y.-J. Doh, J. A. van Dam, A. L. Roest, E. P. A. M. Bakkers, L. P. Kouwenhoven, and S. De Franceschi, Science \textbf{309}, 272 (2005).
\bibitem{xiang}
J. Xiang, A. Vidan, M. Tinkham, R. M. Westervelt, and C. M. Lieber, Nature Nanotech. \textbf{1}, 208 (2006).
\bibitem{pothier}
H. Pothier, S. Gu$\acute{\text{e}}$ron, D. Esteve, and M. H. Devoret, Phys. Rev. Lett. \textbf{73}, 2488 (1994).
\bibitem{courtois}
H. Courtois, Ph. Gandit, and B. Pannetier, Phys. Rev. B \textbf{52}, 1162 (1995).
\bibitem{morpurgo1}
A. F. Morpurgo, B. J. van Wees, T. M. Klapwijk, and G. Borghs, Phys. Rev. Lett. \textbf{79}, 4010 (1997).
\bibitem{giazotto}
F. Giazotto, P. Pingue, F. Beltram, M. Lazzarino, D. Orani, S. Rubini, and A. Franciosi, Phys. Rev. Lett. \textbf{87}, 216808 (2001).
\bibitem{baselmans}
J. J. A. Baselmans, A. F. Morpurgo, B. J. van Wees, and T. M. Klapwijk, Nature \textbf{397}, 43 (1999).
\bibitem{belzig}
W. Belzig, F. K. Wilhelm, C. Bruder, G. Sch$\ddot{\text{o}}$n, and A. D. Zaikin, Superlattices Microstruct. \textbf{25}, 1251 (1999).
\bibitem{belzig2}
W. Belzig, C. Bruder, and G. Sch$\ddot{\text{o}}$n, Phys. Rev. B \textbf{54}, 9443 (1996).
\bibitem{cuevas}
J. C. Cuevas, J. Hammer, J. Kopu, J. K. Viljas, and M. Eschrig, Phys. Rev. B \textbf{73}, 184505 (2006).
\bibitem{gueron}
S. Gu$\acute{\text{e}}$ron, H. Pothier, N. O. Birge, D. Esteve, and M. H. Devoret, Phys. Rev. Lett. \textbf{77}, 3025 (1996).
\bibitem{sueur}
H. le Sueur, P. Joyez, H. Pothier, C. Urbina, and D. Esteve, Phys. Rev. Lett. \textbf{100}, 197002 (2008).
\bibitem{pillet}
J-D. Pillet, C. H. L. Quay, P. Morfin, C. Bena, A. Levy Yeyati, and P. Joyez, Nature Phys. \textbf{6}, 965 (2010).
\bibitem{zhou}
F. Zhou, P. Charlat, B. Spivak, and B. Pannetier, J. Low Temp. Phys. \textbf{110}, 841 (1998).
\bibitem{petrashov}
V. T. Petrashov, V. N. Antonov, P. Delsing, and T. Claeson, Phys. Rev. Lett. \textbf{74}, 5268 (1995).
\bibitem{petrashov2}
V. T. Petrashov, K. G. Chua, K. M. Marshall, R. Sh. Shaikhaidarov,  and J. T. Nicholls, Phys. Rev. Lett. \textbf{95}, 147001 (2005).
\bibitem{ingold} G. L. Ingold and Yu. V. Nazarov, in Single Charge Tunneling, edited by H. Grabert and M. H. Devoret, NATO ASI Series B, Vol. 294 (Plenum Press, New York, 1992), pp. 21–107.
\bibitem{giazotto2}
F. Giazotto, J. T. Peltonen, M. Meschke, and J. P. Pekola, Nature Phys. \textbf{6}, 251 (2010).
\bibitem{clarke}
\textit{The SQUID Handbook}, edited by J. Clarke and A. I. Braginski (Wiley-VCH, Weinheim, 2004).
\bibitem{tinkham}
M. Tinkham, \emph{Introduction to Superconductivity, 2nd Edition} (McGraw-Hill, Inc., New York, 1996).
\bibitem{likharev}
K. K. Likharev, \emph{Dynamics of Josephson Junctions and Circuits} (Gordon and Breach Publishers, Amsterdam, 1986).
\bibitem{usadel}
K. D. Usadel, Phys. Rev. Lett. \textbf{25}, 507 (1970).
\bibitem{wolf}
E. L. Wolf, \emph{Principles of Electron Tunneling Spectroscopy} (Oxford University Press, New York, 1985).
\bibitem{devoret}
M. H. Devoret, D. Esteve, H. Grabert, G.-L. Ingold, H. Pothier, and C. Urbina, Phys. Rev. Lett. \textbf{64}, 1824 (1990).
\bibitem{sns}
H. Courtois, M. Meschke, J. T. Peltonen, and J. P. Pekola, Phys. Rev. Lett. \textbf{101}, 067002 (2008).
\bibitem{dubos}
P. Dubos, H. Courtois, B. Pannetier, F. K. Wilhelm, A. D. Zaikin, and G. Sch$\ddot{\text{o}}$n, Phys. Rev. B {\bf 63}, 064502 (2001). 
\bibitem{heikkila} T. T. Heikkil$\ddot{\text{a}}$, J. S$\ddot{\text{a}}$rkk$\ddot{\text{a}}$, and F. K. Wilhelm, Phys. Rev. B {\bf 66}, 184513 (2002).
\bibitem{rmp}
F. Giazotto, T. T. Heikkil$\ddot{\text{a}}$, A. Luukanen, A. M. Savin, and J. P. Pekola, Rev. Mod. Phys. \textbf{78}, 217 (2006).
\bibitem{HG}
T. T. Heikkil$\ddot{\text{a}}$ and F. Giazotto, Phys. Rev. B \textbf{79}, 094514 (2009). 
\bibitem{timo}
A. V. Timofeev, M. Helle, M. Meschke, M. M$\ddot{\text{o}}$tt$\ddot{\text{o}}$nen, and J. P. Pekola, Phys. Rev. Lett. \textbf{102}, 200801 (2009).


\end{thebibliography}
\end{document}